# On-chip integrated metasystem for spin-dependent multi-channel colour holography

Zhan-Ying Ma,[1,2] Xian-Jin Liu,[1,2] Yu-Qi Peng,[1,2] Da-Sen Zhang,[1,2] Zhen-Zhen Liu,[1,2] Jun-Jun Xiao[1,2,3,\*]

[1] College of Electronic and Information Engineering, Harbin Institute of Technology (Shenzhen), Shenzhen 518055, China

[2] Shenzhen Engineering Laboratory of Aerospace Detection and Imaging, Harbin Institute of Technology (Shenzhen), Shenzhen 518055, China

[3] College of Integrated Circuits, Harbin Institute of Technology (Shenzhen), Shenzhen 518055, China

[\*] Corresponding address: eiexiao@hit.edu.cn

**ABSTRACT：**

**On-chip integrated metasurface driven by in-plane guided waves is of great interests in various light field manipulation applications such as colorful augmented reality and holographic display. However, it remains a challenge to design colorful multichannel holography by a single on-chip metasurface. Here we present metasurfaces integrated on top of guided-wave photonic slab that achieves multi-channel colorful holographic light display. An end-to-end scheme is used to inverse design the metasurface for projecting off-chip preset multiple patterns. Particular examples are presented for customized patterns that were encoded into the metasurface with a single-cell meta-atom, working simultaneously at RGB color channels and for several different diffractive distance, with polarization dependence. Holographic images are generated at 18 independent channels with such a single-cell metasurface. The proposed design scheme is easy to implement and the resulting device is viable to fabrication, promising a plenty of applications in nanophotonics.**

Augmented reality (AR) is a display technology that can integrate virtual images with the real scenes, enhancing the interactivity between the display technology and the surrounding environment [1-8]. On the other hand, metasurface is an artificial structure that can modulate the phase, amplitude, polarization, and spectral characteristics of light waves based on the design of the surface nanostructures. It can work for beam deflection [9-11], metalens [12-13], and holographic display [14-15]. Integrating a metasurface onto an optical waveguide can bridging the manipulation of on-chip guided waves and free space light. This contributes to the development of chip level photonic devices and on-chip photonic integration [16-18]. On-chip integrated metasurfaces driven by guided waves have shown great potentials [19-23], which also provide more possibilities for holographic display due to their compactness and comprehensive manipulation of light waves. In order to maximize the functionalities of AR holographic devices, it is desirable to engineer the multi-color and multi-channel image projection, yet efficient inverse design for that remains unexplored.

The design of on-chip metasurface structures for multiple channel functionalities posts great challenge for the forward methods that starts by selecting a targeted phase distribution [27]. Here we propose an end-to-end optimization method for the on-chip metasurface design. This method connects the target light field with the metasurface geometry and finds the optimal solution of the system by iteratively updating the structural parameters. We show that it is possible to have multicolor spin-dependent multiplexing holographic image projection with single-cell meta-atoms.

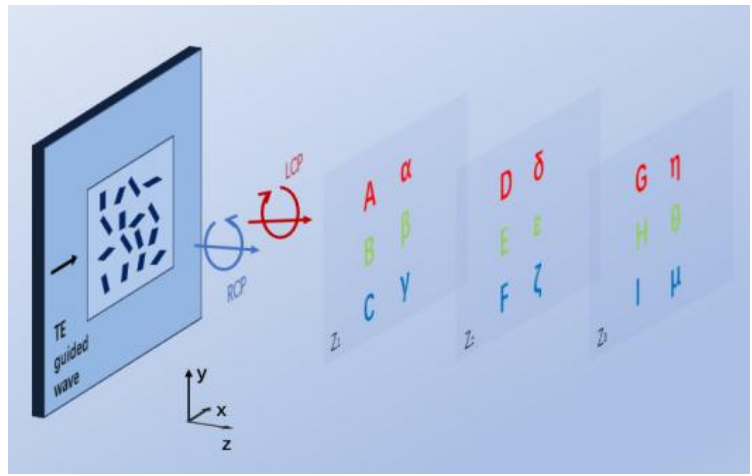

**Fig.1.** Schematic diagram of on-chip integrated system for multi-channel pattern projection. On-chip propagating guided waves interact with the metasurface and are scattered light out to free space with polarization-dependent wavefronts that are different for multiple operation frequency, forming targeted patterns on the different plane.

Figure **1** schematically shows our proposed system for 3 color channel, 3 diffraction distance channel, and 2 polarization channel and may amount up to totally 18 channels. The device consists of a metasurface integrated on a waveguide (See Supplement 1). Light propagating inside the waveguide interacts with the meta-atoms and couples off the chip. We encoded arbitrarily customized holographic images into the metasurface on the waveguide, achieving the extraction of guided waves transmitted within the

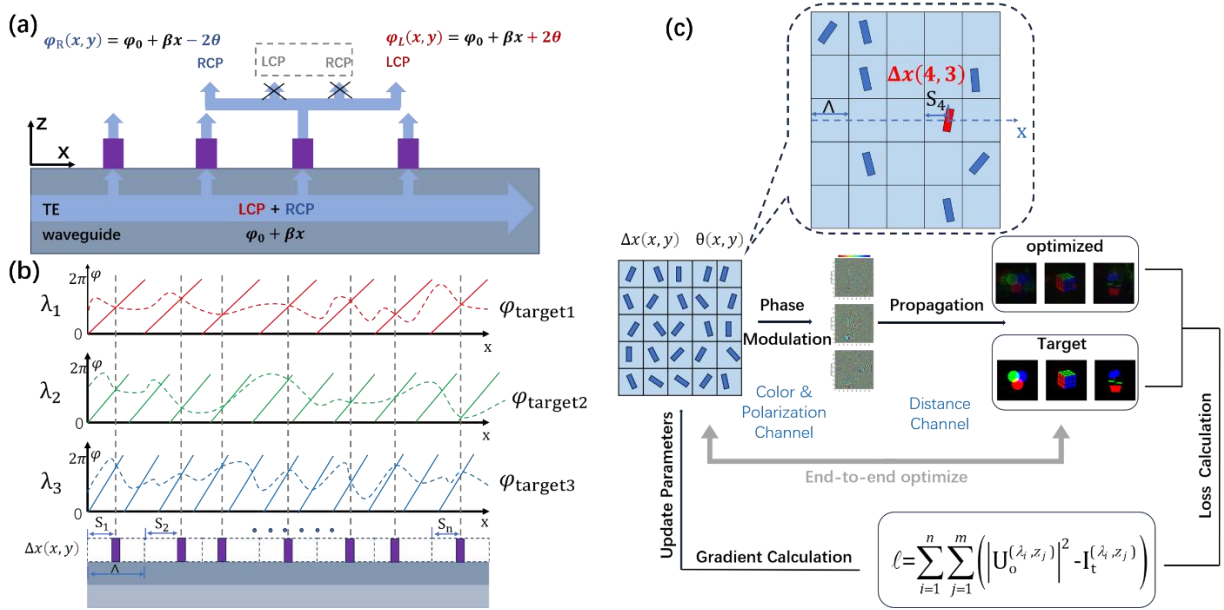

**Fig.2.** Illustration of the working principle of metasurface excited by guided wave and the end-to-end inverse design pipeline. **(a)** Traveling TE mode can be decomposed into the components of LCP and RCP waves respectively, both of which are coupled out and independently manipulated by the elaborately designed geometric metasurface with different phase profiles for function multiplexing. **(b)** Schematic diagram of the relationship between phase modulation of each wavelength channel and the meta-atom displacement (along the TE wave propagation direction) when only the propagation accumulation phase is considered. The dashed curve represents a final required phase and the straight solid lines are wrapped phase. The intersection point between the dashed line and the straight line represents the optimally searched phase modulation for each wavelength channel and the meta-atom position. **(c)** Schematic diagram of the end-to-end optimization method for the design of metasurface structures that works at multiple wavelength $\lambda_i$ and target plane $z_j$. This method connects the target light fields with the metasurface structure and achieves the optimal solution of the system by continuously updating the structural parameters $\Delta x(x,y)$ and $\theta(x,y)$. In our design, guided waves propagate along the $x$ direction. We do not consider the displacement of meta-atoms along the $y$ direction and set it to a constant value (half of period length). $\Delta x(x,y)$ represents the displacement along the $x$ direction within a period of $\Lambda$ of the meta-atom in the x-th column and y-th row of the on-chip integrated metasurface. The same principle goes for $\theta(x,y)$.

waveguide and generating visible holographic images on different planes in the free space. We note that previous work on guided-wave driven metasurface mostly focus on single frequency operation [21,25,28], in sharp contrast to our study here. Most importantly, by taking advantage of the linear polarization characteristic of the guided wave mode (for example, here TE mode) and the geometric phase by anisotropic meta-atoms, we are able to separate the spin-dependent left-circular polarization (LCP) and the right-circular polarization (RCP) light projection.

Figure **2(a)** shows the geometric metasurface which is composed of anisotropic meta-atoms with identical shape and varying in-plane orientation. The meta-atoms can be directly driven by an eigen-TE guided mode [24] (more details on waveguide geometry and the guided mode dispersion are in Supplement I). The on-chip guided waves are perturbed by meta-atoms to reconstruct the wavefront of scattered waves that propagate upward into free space. Along the $x$ direction (assume guided wave propagating along the $x$ direction), the phase distribution of the extracted waves can be expressed as

$$\varphi(x) = \varphi_0 + \beta x \tag{1}$$

where $\varphi_0$ is the initial phase of the incident light, $\beta$ is the propagation constant and therefore $\varphi_0 + \beta x$ is the propagation accumulated phase. Besides, a travelling TE guided wave can be regarded as a superposition of LCP and RCP waves. Therefore, rotation of a meta-atom with in plane orientation angle $\theta$ shall induce a Pancharatnam-Berry (PB) phase $2\theta$ for the scattered LCP wave, followed by an opposite phase shift $-2\theta$ for the RCP wave [24]. Therefore, when considering both the in-grid relative displacement and orientation of the meta-atoms, the scattering wave phase would be determined by both the propagation accumulated phase and the PB phase, which reads

$$\varphi_R^{(\lambda)}(x,\theta) = \varphi_0 + \beta_\lambda x - 2\theta \tag{2}$$

$$\varphi_L^{(\lambda)}(x,\theta) = \varphi_0 + \beta_\lambda x + 2\theta \tag{3}$$

Note that for color dependent light-field projection, we must consider the wavelength dependence of the propagation constant $\beta_\lambda$ (see Supplement I). Now we can obtain the required target phase by encoding the displacement and orientation of the meta-atoms.

For design task involving single wavelength $\lambda_1$, considering single-cell structure and for ease of calculation, the metasurface pixel period can be selected as

$$\Lambda = \frac{2\pi}{\beta_{\lambda_1}} = \frac{\lambda_1}{n_{eff}} \tag{4}$$

where $n_{eff}$ is the effective reflective index of the waveguide mode operating at $\lambda_1$. Hence, the detour phase of the light extracted for the nth meta-atom is derived as

$$\Delta\varphi_n = \frac{2\pi}{\Lambda} \cdot S_n \tag{5}$$

where $s_n$ is the n-th meta-atom displacement along the $x$ direction within a period of $\Lambda$. For each target holographic display image, the displacement of the meta-atom in each pixel unit is determined by equation (5), which means that for any customized pattern, we can encode the displacement information of the meta-atom of the on-chip metasurface through the detour phase.

To drive the inverse design loop, one must first obtain a phase $\varphi_{target}$ based on the target pattern. The acquisition of this phase can be achieved through the GS (Gerchberg–Saxton) algorithm [26], and then the target phase can be mapped to the displacement $s_n$ of the meta-atom in each pixel. This approach can conveniently meet the requirements of one target pattern, however, it is difficult to achieve multiplexing of more channels because the structure of the on-chip metasurface has already been encoded and determined by the specific customized pattern. We note that there were studies that have achieved multiplexing of channels in two orthogonal directions, obtaining two different holographic projections from the orthogonal directions [21,28]. However, in the above scheme, the on-chip metasurface is designed to operate at a single wavelength in both the two guided wave incoming directions. It is very important to achieve the operation of multiple wavelength channels when light is incident from the same direction, as well.

In order to achieve more wavelength channel multiplexing, we proposed an end-to-end optimization method to design the structure of on-chip metasurface. We believe this represents the most significant contribution of our study. The method is actually universal and can be widely adopted.

As a concrete example, let us take the representative wavelength $\lambda_1, \lambda_2$ and $\lambda_3$ for RGB color channels, respectively. The displacement of meta-atom of the on-chip metasurface and its corresponding phase in the RGB color channel are shown in Figure **2(b)**. Taking the red channel as the reference, according to Equation (1), the slope of the red line in Figure **2(b)** is $\beta_{\lambda_1}$. According to Equation (4), for guided wave goes in the $x$ direction for length $\Lambda$, the accumulated propagation phase is $2\pi$ and could be wrapped up. It is then nature to choose the metasurface pixel grid pitch as $\Lambda$ [21,27,28]. However, in the case of multicolor channels, due to the different propagation constant $\beta_\lambda$, the pixel pitch $\Lambda$ cannot satisfy Equation (4) to meet the requirements of three color channels simultaneously. In view that the phase generated by the GS algorithm must be discretized, here we recognize that the division of pixel periods is actually not necessary and it does not strictly require a periodic sampling of the phase. In other words, some points could be missed but the impact on the quality of the final holographic projection could still be acceptable. As Figure **2(b)** illustrates, the meta-atom position setting doesn't intersect each of the wrapped propagating phase for $\lambda_2, \lambda_3$. We note that this is somehow like the spatial multiplexed color filter for metasurface [29].

With the above understanding, we then inverse design the structure for multicolor operation. Figure **2(c)** shows the end-to-end inverse design pipeline. One first selects an initial displacement map $\Delta x$, and then calculates the output field by the Raleigh-Sommerfeld (RS) diffraction formula, with the detour phase given by Equation (5). The loss function is defined as

$$\ell = \sum_{i=1}^{n}\sum_{j=1}^{m}\left(\left|U_o^{(\lambda_i,z_j)}\right|^2 - I_t^{(\lambda_i,z_j)}\right) \tag{6}$$

where $U_0$ is the output from the theoretical model [Eqs. (1)-(5)] and the RS formula] and $I_t$ is the target pattern. Note that the loss function dependents on the meta-atom displacement map $\Delta x(x,y)$ and the rotation angle map $\theta(x,y)$ which are both updated iteratively to minimize $\ell$ by the gradient decent approach.

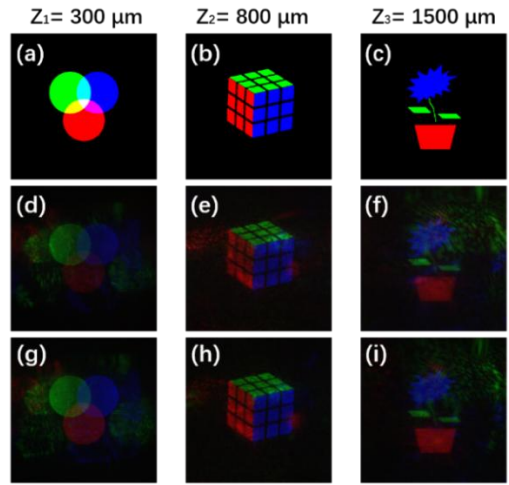

**Fig.3.** Holographic images of muti-channel multiplexing metasurface inverse designed by the end-to-end scheme. The holographic images are reproduced at 3 distances ($z_1 = 300$ μm, $z_2 = 800$ μm and $z_3 = 1500$ μm), each with three wavelength channels ($\lambda_1 = 450$ nm, $\lambda_2 = 532$ nm and $\lambda_3 = 635$ nm). **(a)-(c)** the target images. **(d)-(f)** Results by considering only the propagation accumulation phase, i.e., using only the meta-atoms displacement as the design parameters. **(g)-(i)** Results by considering both the propagation accumulation phase and the PB phase, phase modulation design was carried out specifically for the RCP wave.

As a comparison, we also considered the case where only the displacement map $\Delta x(x,y)$ of the meta-atoms is used as the design space. Figure **3** shows results for three wavelength channels ($\lambda_1 = 450$ nm, $\lambda_2 = 532$ nm and $\lambda_3 = 635$ nm) and three distance channels ($z_1 = 300$ μm, $z_2 = 800$ μm and $z_3 = 1500$ μm), totally 9 channels for holographic projection multiplexing. We have selected the three distances randomly and there is in principle no particular limitation on the number of imaging planes, depending on the signal-to-noise constrain. One can therefore generate a quasicontinium 3D holographic object [30].Figures **3(a)-3(c)** are the target patterns. Figures **3(d)-3(f)** are the optimized results considering only the displacement map, while Figures **3(g)-3(i)** are the optimized results for design space containing both $\Delta x(x,y)$ and $\theta(x,y)$. The detailed optimization convergence and the final design parameters can be found in Supplement I. The results clearly show that the expansion of the parameter space has improved final results.

In subsequent examples, we shall consider both the displacement and orientation of the meta-atoms as the design space. It should be pointed out that although the results of Figures **3(g)-3(i)** look more clear than those in Figures **3(d)-3(f)**, the actual intensity of Figures **3(g)-3(j)** should be lower than those in Figures **3(d)-3(f)** simply because of polarization control is needed to get the former.

In the above example, we have achieved 9 channel multiplexing for the scattered RCP waves. In fact, phase shifts ($\pm 2\theta$, shown in Equation (2) and Equation (3)) introduced by orientation can be simultaneously manipulated to achieve polarization demultiplexing, which means the phase modulation to the scattered LCP and RCP waves can be considered simultaneously to generate both RCP and LCP target light field [24]. By considering both 2 polarizations, the multiplexing channel number doubles to 18. We take 9 Alphabetic letters as RCP target holographic images and 9 Greek letters as LCP target holographic images. Figure **4** shows the optimized results. More detailed optimization convergence and the final design parameters can be found in Supplement I. The target holographic images with different polarizations are successfully reproduced at different distances with different colors through our design method.

To demonstrate the feasibility of our proposed method, limited by computational resources, we reduced the number of meta-atoms and conducted 3D full-wave simulation by FDTD. More details can be found in Supplement I.

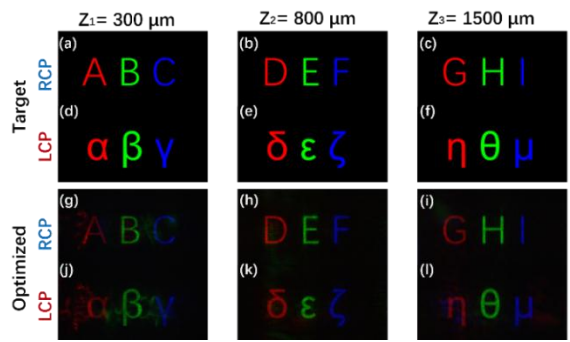

**Fig.4.** Optimized results for the on-chip metasurface working in 18 channels that projects color Alphabetic letters and Greek letters. **(a)-(c)** The target images of Alphabetic letters with RCP, which are reproduced at 3 different distances, with 3 colors in each plane. **(d)-(f)** The target images of Greek letters with LCP, which are reproduced at 3 different distances, with 3 colors in each plane. **(g)-(i)** The optimized results of RCP waves. **(j)-(l)** The optimized results of LCP waves.

In summary, we propose an inverse design scheme for on-chip integrated metasurfaces, which can achieve multi-channel colour holographic display in the visible. By combining the wavelength, diffractive distance, and circular polarization, we can achieve holographic projection of images equivalent to 18 channels, effectively enriching the capabilities of the on-chip integrated metasurface. This system is small, compact, and has comprehensive functions, which can be widely used in wearable devices such as AR devices and for other nanophotonic devices.

**Funding**. National Natural Science Foundation of China (No. 62375064), Shenzhen Science and Technology Program (No. JCYJ20210324132416040), Guangdong Provincial Nature Science Foundation (No. 2022A1515011488), and the National Key Research and Development Program of China (No. 2022YFB3603204), Guangdong Basic and Applied Basic Research Foundation (No. 2023A1515110572)

**Disclosures**. The authors declare no conflicts of interest.

**Data Availability**. Data underlying the results presented in this paper are not publicly available at this time but may be obtained from the authors upon reasonable request.

**Supplemental document.** See Supplement for supporting content.

# Supplementary Information

## 1. Mechanism of the designed on-chip integrated metasystem for multiplexing.

The on-chip integrated metasystem we designed has a three-layer structure (see **Figure S1(a)**). An amorphous silicon (α-Si) is placed on the upper layer of the planar waveguide of Si3N4 to couple the propagating light inside the waveguide into free space. The bottom layer is SiO2 serving as the bottom substrate. In order to avoid the influence of the close distance between two meta-atoms on the coupling of guided waves into free space, we limit the distance between two meta-atoms (the distance between the centers of adjacent meta-atoms) to be no less than 150 nm during the design(see **Figure S1(c)**). The designed on-chip metasurface has 900×900 pixels with the pixel period $\Lambda = 400$ nm.

**Figure S1(c)** is the guided mode dispersion showing dependence of the effective refractive index $n_{eff}$ on the wavelength $\lambda$ in absence of the metasurface. For different wavelengths, the propagation constant $\beta_i$ is different, which leads to different propagation accumulation phases for same propagation distance. This is the basis for achieving color channel multiplexing. The modal fields are shown in **Figure S2**.

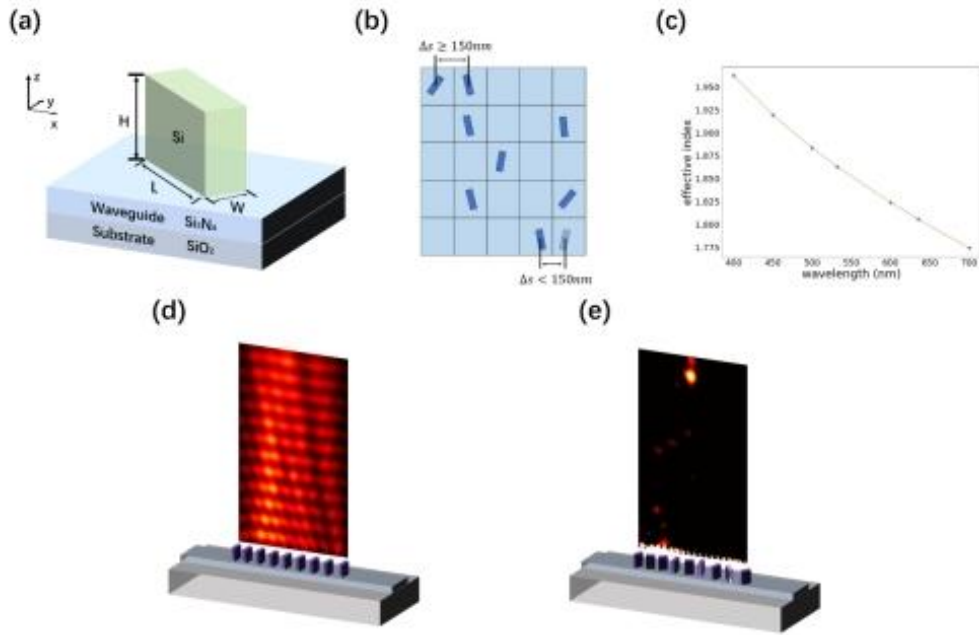

**Fig. S1**. **(a)** Schematic of the α-Si meta-atom with the height of H = 380 nm, length of L =180 nm and width of W = 90 nm. **(b)** The distance between two adjacent meta-atoms needs to be no less than 150 nm. **(c)** The guided mode dispersion showing the relationship between the effective refractive index and the wavelength. **(d)** FDTD simulation results for the case that the displacement of meta-atoms are set in a way to generate vertically out-going plane wave purely by the detour phase. **(e)** FDTD simulation results for the design that uses the rotation angle of the meta-atoms in (d) and generate additional PB phase to focus the out-going wave to a point.

As a proof-of-concept demonstration for our model and to verify the generation of detour phase and PB phase, we position the meta-atoms on a strip waveguide and generate vertical scattering wave emission from the metasurface by encoding the displacement of the meta-atoms. We further encode the rotation angle of the meta-atoms to to create a 1D on-chip metalens that focuses the scattered wave into a point in the free space. Note that here 80 meta-atoms are placed on the strip waveguide and the focus length is set to 8μm. FDTD simulation is conducted for the designed metasurface and resutls as shown in **Figures S1(d)-S1(e)**. It is seen that the FDTD results are as expected.

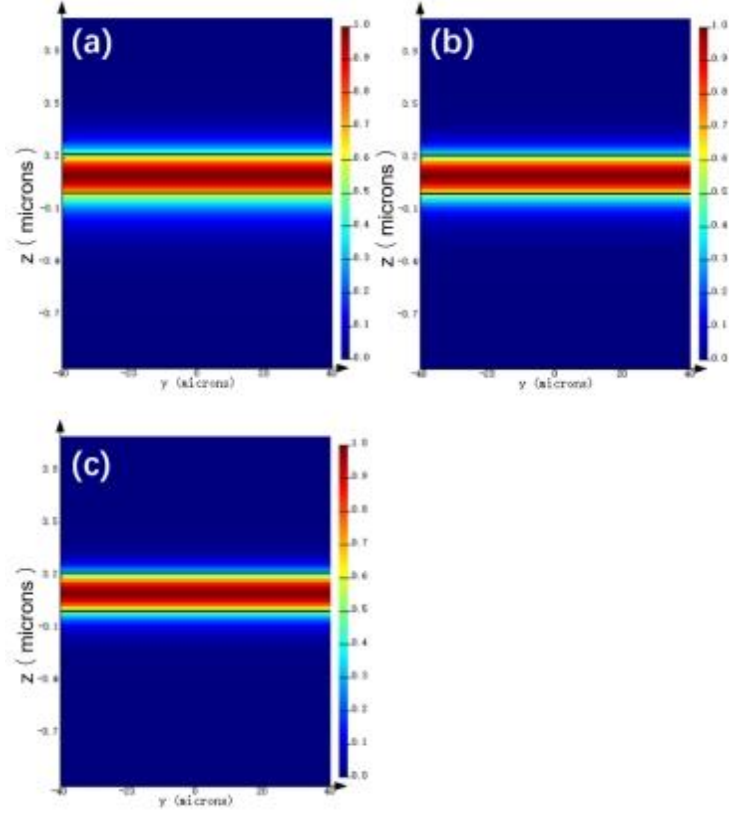

**Fig.S2.** $E_y$ component of modal field for (a) $\lambda_1$ = 450 nm  (b) $\lambda_2$ = 532 nm (c) $\lambda_3$ = 635 nm.

## 2. The final design parameters and phase modulation of the designed metasurface

The first example takes 9 pictures as target holographic patterns. The superposed 3 images in the same plane with different colors form holographic image like magic cube, etc. The optimized results shown in **Figure S3** are obtained by considering only the propagation accumulation phase, thus only the displacement distribution $\Delta x(x,y)$ of the meta-atoms (shown in **Figure S3(a)**) is presented. The optimized structure has different phase modulation for three colors (shown in **Figures S3(b)-S3(d)**), combined with three distance channels in space, achieving multiplexing of 9 channels.

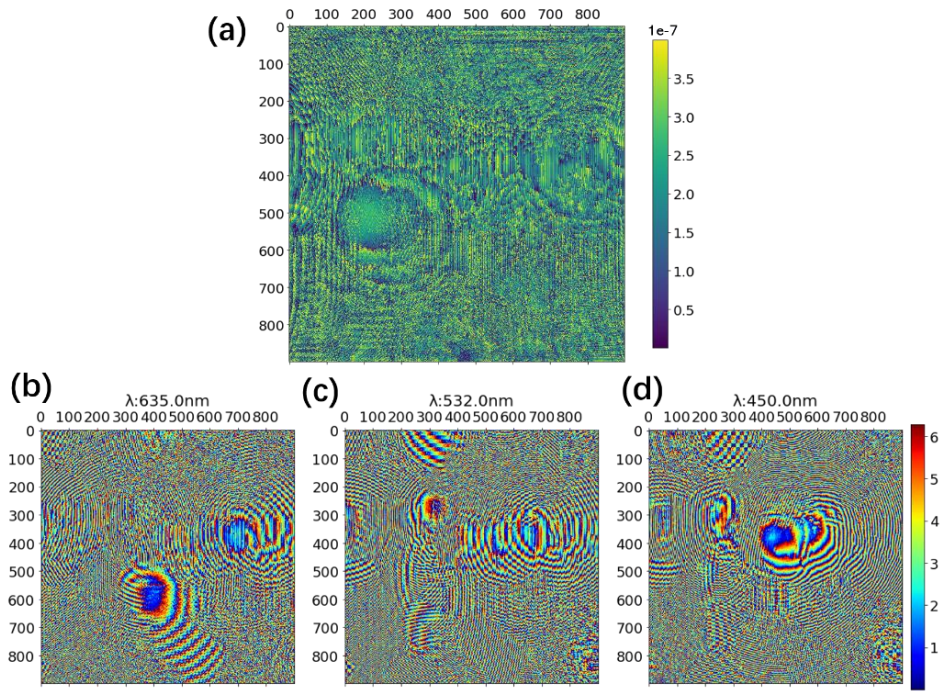

**Fig.S3.** The optimized design space map. Results were obtained by considering only the propagation accumulation phase, i.e. using only the meta-atoms displacement as the design parameters. **(a)** The displacement distribution Δx(x,y) of the meta-atoms. **(b)-(d)** Phase modulation of the optimized on-chip metasurface structure for three operating wavelength.

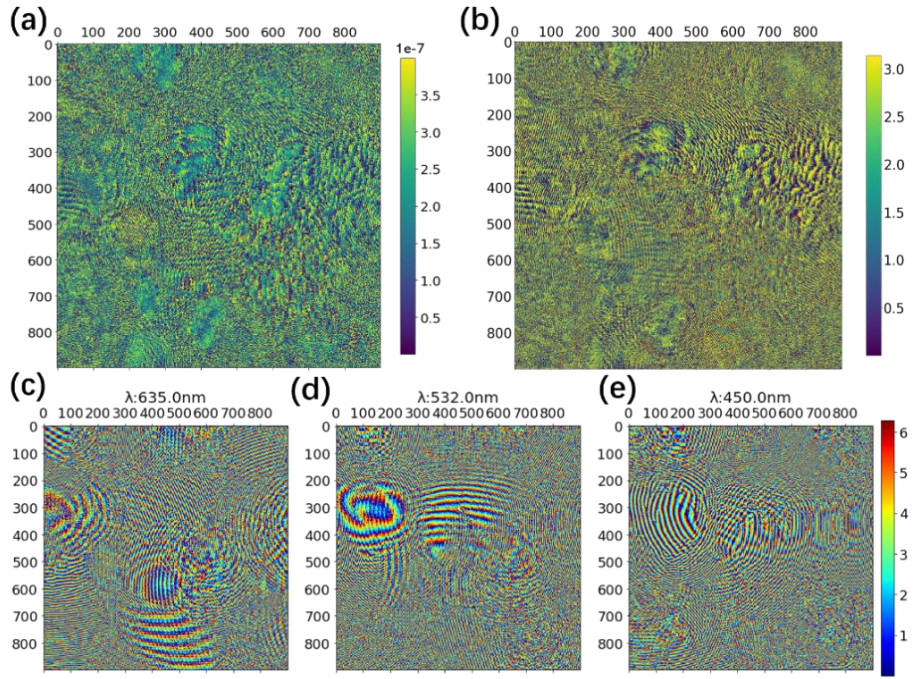

**Fig.S4.** The optimized design space map. Results were obtained by considering both the propagation accumulation phase and the PB phase, and the phase modulation design was carried out specifically for the RCP wave. **(a)** The displacement distribution Δx(x,y) of the meta-atoms. **(b)** The orientation distribution θ(x,y) of the meta-atoms. **(c)-(e)** Phase modulation to the scattered RCP waves of the optimized on-chip metasurface for three different wavelength.

As a comparison, we expand the parameter space, considering both the propagation accumulation phase and the PB phase. Keep other conditions unchanged and phase modulation design was carried out specifically for the RCP wave. The optimized results are shown in **Figure S4.** Phase modulation is determined by both The displacement distribution $\Delta x(x,y)$ of the meta-atoms and the orientation distribution $\theta(x,y)$ of the meta-atoms.

The next example takes 9 Alphabetic letters as RCP target holographic images and 9 Greek letters as LCP target holographic images. By considering both the RCP and LCP polarizations, the multiplexing channel number doubles (up to 18). Detailed optimized results are shown in **Figure S5.**

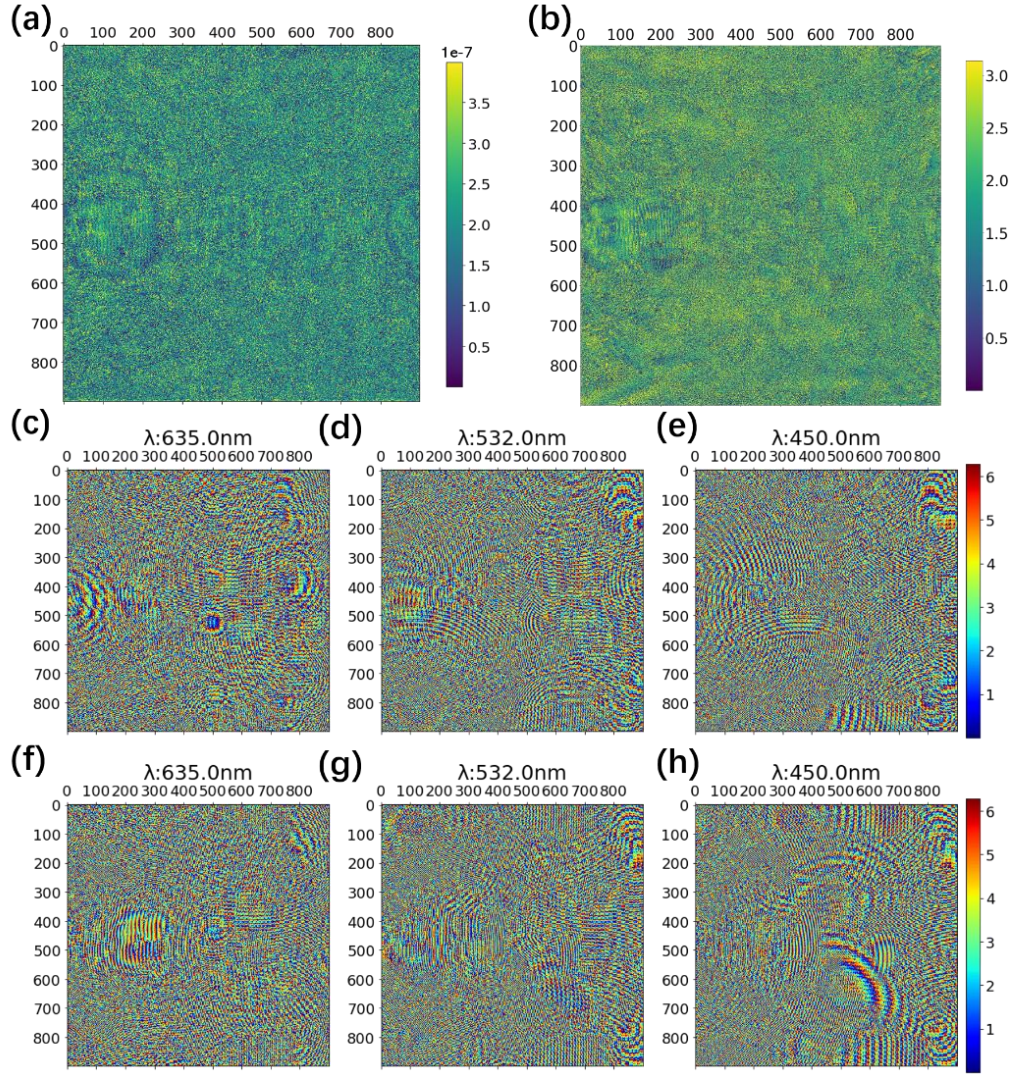

**Fig.S5.** The optimized design space map. Results were obtained by considering both the propagation accumulation phase and the PB phase, phase modulation design was carried out for both the RCP wave and the LCP wave. **(a)** The displacement distribution $\Delta x(x,y)$ of the meta-atoms. **(b)** The orientation distribution $\theta(x,y)$ of the meta-atoms. **(c)-(e)** Phase modulation to the scattered RCP waves of the optimized on-chip metasurface structure for different wavelength. **(f)-(h)** Phase modulation to the scattered LCP waves of the optimized on-chip metasurface structure for different wavelength.

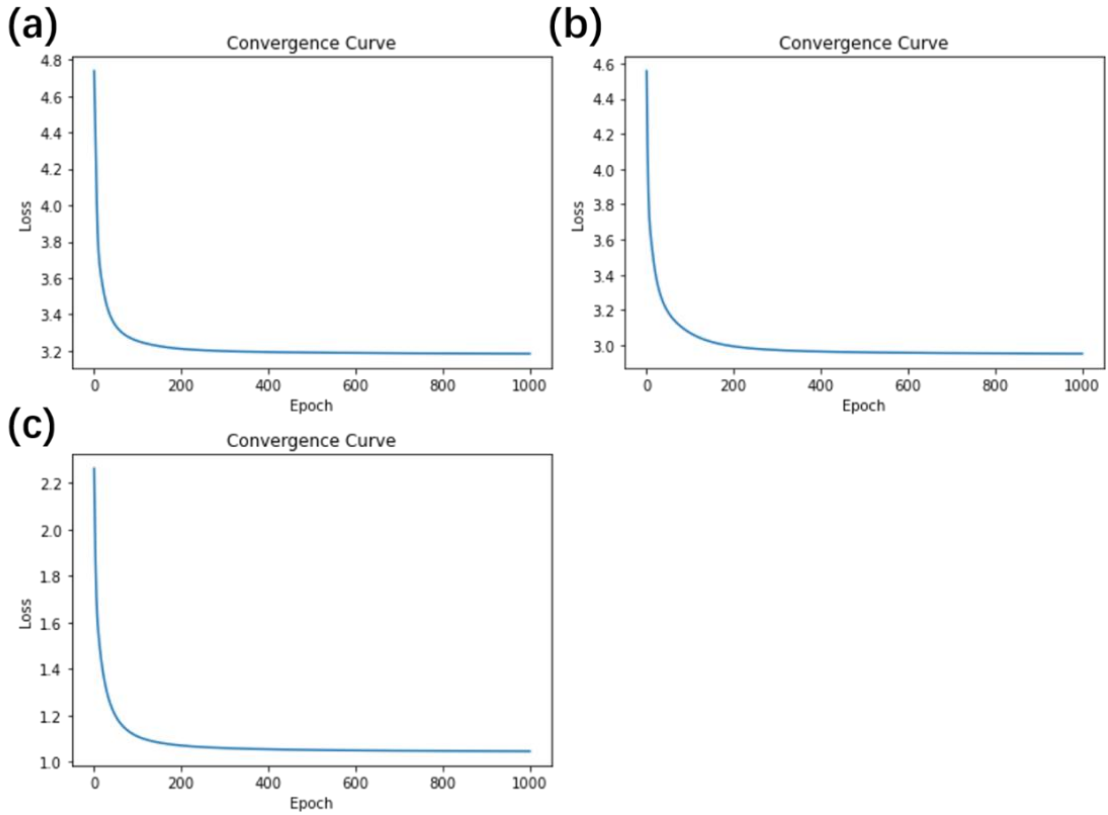

**Fig.S6.** The optimization convergence curve. **(a)** Loss convergence during the optimization process considering only the propagation accumulation phase **(b)** Loss convergence during the optimization process considering both the propagation accumulation phase and the PB phase, only designed for the RCP wave **(c)** Loss convergence during the optimization process considering both the propagation accumulation phase and the PB phase, designed for both the RCP and LCP waves.

The iterative optimization converges for all mentioned examples and the results are shown in **Figure S6.** For the same target holographic patterns, the expansion of the parameter space does indeed reduce loss and improve the final optimization effect. It should be pointed out that the lower loss in **Figure S6(c)** is due to our selection of different target holographic patterns, which have smaller pixel occupancy.

**3. 3D full-wave simulation by FDTD**

To ensure the holographic effect, it is necessary to adopt a metasurface with fairly large number of meta-atoms and therefore a pretty large size on-chip metasurface, which posts great challenges on 3D full-wave simulation by FDTD. In order to reduce the computing resources required yet still showcase our proposed end-to-end reverse design method, we consider a system of up to 80×80 meta-atoms for the design loop and obtained the optimize structures for full-wave FDTD simulation. All holograms are designed to be reproduced at the plane of 20μm. The optimized results and 3D full-wave simulation results by FDTD are shown in the **Figure S7**. The results clearly demonstrate the feasibility of our proposed method and also reflect the fact that the reduction in the number of meta-atoms and the size of the metasurface have an impact on the hologram.

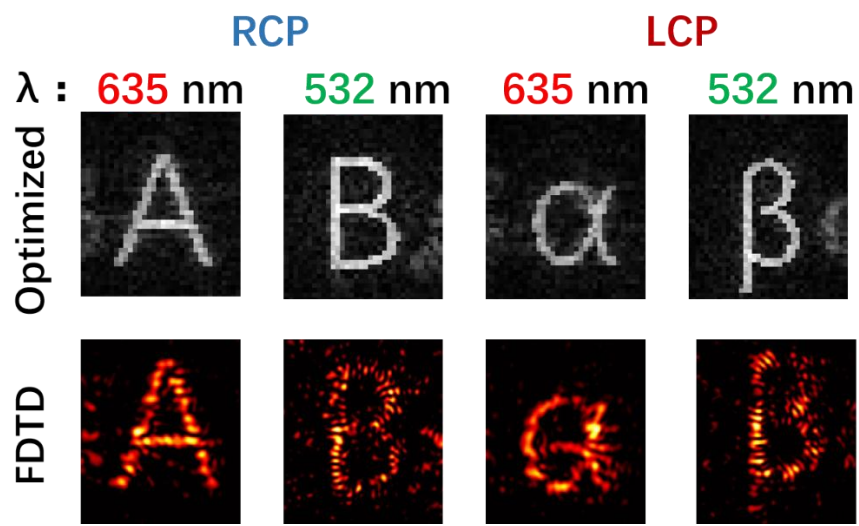

**Fig. S7.** Optimized results for size-reduced on-chip metasurface and the holographic results by FDTD simulation.